\documentclass{elsart}

 \usepackage{graphicx}

\usepackage{amssymb}

\begin{document}

\begin{frontmatter}



\title{New approach for network monitoring and intrusion detection}


\author{Vladimir Gudkov\corauthref{cor}}
 \ead{gudkov@sc.edu}
 \corauth[cor]{Corresponding author.}
\author{and}
\author{Joseph E. Johnson }
\ead{jjonson@sc.edu}
\address{Department of Physics and Astronomy \\
University of South Carolina \\ Columbia, SC 29208 }

\begin{abstract}
The approach for a network behavior description  in terms of
numerical time-dependant functions of the protocol parameters is
suggested. This provides a basis for application of methods of
mathematical and theoretical physics for information flow analysis
on network and for extraction of patterns of typical
 network behavior. The
information traffic can be described as a trajectory in
multi-dimensional parameter-time space with dimension about 10-12.
Based on this study some algorithms for the proposed intrusion
detection system  are discussed.
\end{abstract}


\end{frontmatter}

\section{Introduction}
\label{inr}

Information systems are threatened today as never before. Evolving
network technologies have provided powerful information-transfer
applications that have increased our reliance on computerized
information systems at the same time that public access to the
internet has increased the number and sophistication of those who
seek to compromise these systems.  The diversity of existing
software are hardware devices make the problem of the network the
protection very more difficult.

There are many approaches to study the attack tolerance of
networks. These approaches usually include the study of dependance
of the network stability on the network complexity and topology
(see, for example \cite{cnet1,cnet2} and references therein), and
signature-based analysis technique and statistical analysis and
modelling of network traffic (see, for example
\cite{sig1,st1,st2,st3}). Some methods to study spatial traffic
flows traversing the network \cite{spflow} and correlation
functions of irregular sequences of numbers occurring in the
operation of computer networks \cite{timetr} have been proposed
recently.

In this paper we will not discuss properties related to network
structure and topology, but rather concentrate on the information
flow on the network. The main difference between this paper and
the existing methods is that we are interested in general
properties of information flow on a network. We suggest a new
approach for network monitoring and intrusion detection based on
complete network monitoring. The term ``complete'' will be
clarified later. Since, the process of information exchange is
extremely complicated for detailed analysis, we apply some methods
for analysis of complex systems in physics to the information flow
on networks. This gives us the powerful tools for the analysis and
provides a guideline for the application of the obtained result
for practical purposes.

We will start from a careful analysis of information exchange on
networks to choose the appropriate method of the information flow
description.  It will give us the basis for network simulations in
terms of mathematical description of information exchange
processes. Also, the understanding of information flow on the
network is necessary for real time network monitoring and for
solutions of difficult problems of intrusion detection\cite{indet}
 in real time.

\section{Information flow description}

To describe the information flow on a network, we work on the
level of packets exchanging between computers.  The structure of
the packets and their sizes vary from each to another and depend
on the process. In general each packet consists from a header and
attached (encapsulated) data. Since the data part does not affect
the packet propagation through the network we consider in this
paper only information included in headers. We recall that the
header consists of encapsulated protocols related to different
layers of communications: from a link layer to an  application
layer. The information contained in the headers controls all
network traffic. To extract this information one can use  tcpdump
utilities, developed with the standard of LBNL's Network Research
Group \cite{tcpd}.
 This information is used to analyze network traffic, to
find a signature of abnormal network behavior and to detect
possible intrusions.

The important difference of the proposed approach from
traditionally used methods is the presentation of information
contained in headers in terms of well-defined mathematical
functions. To do that we have developed software to read binary
tcpdump files and to represent all protocol parameters as
corresponding time-dependent functions. This gives us the
opportunity to analyze complete information (or a chosen fraction
of complete information: a combination of some parameters) for a
chosen time and time window. The ability to vary the time window
for the analysis is important since it gives the possibility to
extract different scales in the time dependance of the system.
Different scales have different sensitivities for particular modes
of the system development and, for an example, could be sensitive
to different methods of intrusions.

It should be noted that not all parameters are reliable for the
description of the information flow. The protocol parameters for
host-to-host communication could be divided into two separate
groups\cite{gj1} in respect to the preservation or changing of
their values during the packet propagation through the network. We
call these two groups of parameters as dynamic and static. The
dynamic parameters may be changed during the packet propagation
through the network (internet). For example, a ``physical''
address of a computer, which is  the MAC parameter of the Ethernet
protocol, is a dynamic parameter because it can be changed if the
packet has been re-directed by a router. On the other hand, the
source IP address is an example of static parameter because it
does not change its value during the packet propagation. To
describe the information flow, we use only static parameters
since they may carry intrinsic properties of the information flow
neglecting the network (internet) structure.

Now, using packets as a fundamental objects for information
exchange on network and being able to describe them in terms of
functions of time for static parameters, we can apply methods,
developed in physics and applied mathematics to study complex
dynamical systems, for the network traffic analysis.

To demonstrate the power of these methods we recall an important
result for future applications which has been obtained in the
framework of the presented approach. It was shown\cite{gj1} that
to describe the information flow on network one can use a small
number (10 - 12) of parameters. In other words, the dimension of
the information flow space is less or equal to 12 and the
information flow properties are independent on network structure,
size and topology.  To estimate the dimension of the information
flow on the network one can apply the algorithm  for analysis
observed chaotic data in physical systems suggested in
paper\cite{abar1}(see also ref.\cite{chaos}and references
therein). The main idea is related to the fact that any dynamical
system with dimension $N$ can be described by the set of $N$
differential equations of the second order in configuration space
or by the set of $2N$ differential equations of first order in
phase space.

Assuming that the information flow can be described in terms of
ordinary differential equations (or by discrete-time evolution
rules) for some unknown functions  in a (parametric) phase space,
one can analyze a time dependance of given scalar parameter $s(t)$
which is related to the system dynamics (see, for details
ref.\cite{gj1}). Then one can build $d$-dimensional vectors from
the variable $s$ as
\begin{equation}
y^d(n)=[s(n),s(n+T),s(n+2T),\ldots , s(n+T(d-1))] \label{yvec}
\end{equation}
 at equal-distant time intervals $T$:
 $s(t) \rightarrow s(T\cdot n) \equiv s(n)$, where $n$ is
integer number to  numerate $s$ values at different time. Now one
can calculate a number of nearest neighbors in the vicinity of
each point in the vector space and plot the dependance of the
number of false nearest neighbors(FNN) as a function of time. The
false nearest neighbors for the $d$-dimensional space are
neighbors which move far away when we increase dimension from $d$
to $d+1$. This algorithm could be illustrated by the simple
example for a two-dimensional circle. If we project the circle on
one-dimensional space, we get interval with two degenerated points
along the projection axis. Increasing the dimension by 1 we come
to the original two-dimensional circle without the degeneracy.
Thus, the degenerated points in 1-dimension which have moved to
the opposite sides of the circle in 2-dimensions could be called a
false nearest neighbors (FNN). Unfolding the space further, to
3-dimension, and further we will not get a false nearest neighbors
anymore since a higher dimensional space covers the
two-dimensional space to which the circle belongs.

\begin{figure}
\includegraphics{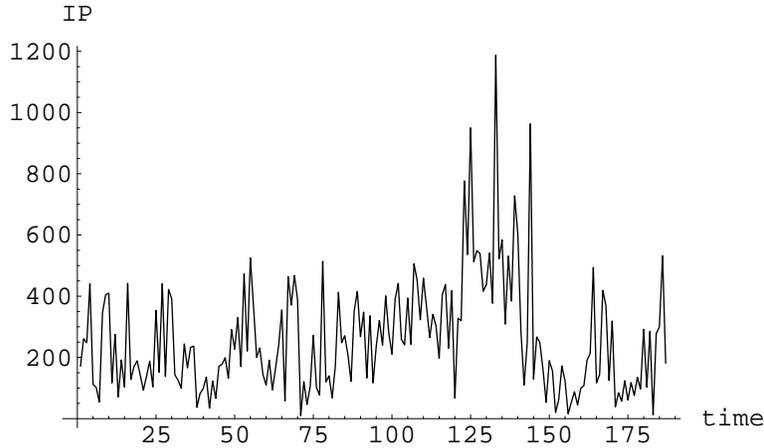}
\caption{Protocol type ID in the IP protocol as a function of time
(in $\tau = 5 sec$ units).} \label{fig:ip}
\end{figure}

\begin{figure}
\includegraphics{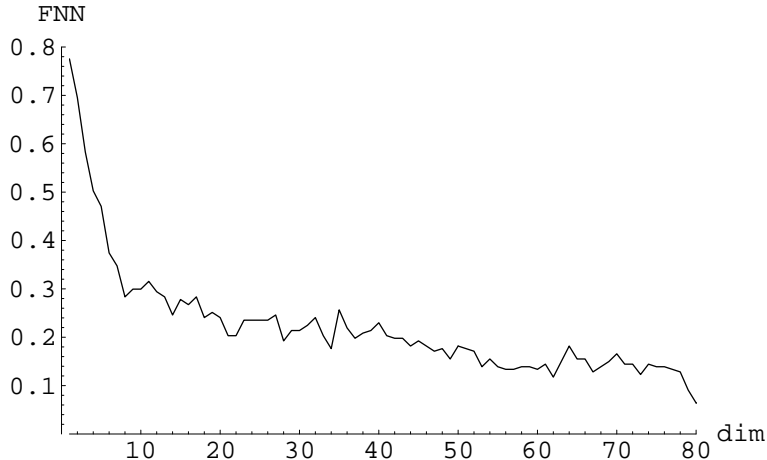}
\caption{Relative number of false nearest neighbors as a function
of dimension of unfolded space.} \label{fig:fnn}
\end{figure}

The typical behavior a scalar parameter and corresponding FNN plot
are shown on Figs. (\ref{fig:ip}) and (\ref{fig:fnn}).
From the last plot one can see that the number of false nearest
neighbors rapidly  decreases  up to about 10 or 12 dimensions.
After that it shows a slow, if any dependance at all,  on the
dimension. This reflects the fact that by increasing the dimension
$d$ step-by-step, the number of false nearest neighbors (FNN),
which come  due to projection of far away parts of the trajectory
in higher dimensional space, is decreasing up to the level
restricted by the system noise since the noise has infinite
dimension.

The analysis of information flow on network has been done in
paper\cite{gj1}  for a wide set of different network
configurations and it was shown that it's is about 10 or 12. It
means one needs not more than 12 independent parameters for a
complete description of the information flow and that its dynamics
can be described as a trajectory in a phase space with the
dimension about 10 - 12. It is also important that this dimension
does not depend on the network topology, its size, and the
operating systems involved in the network. Therefore, this
characteristic is an universal and may be applied for any network.

\begin{figure}
\includegraphics{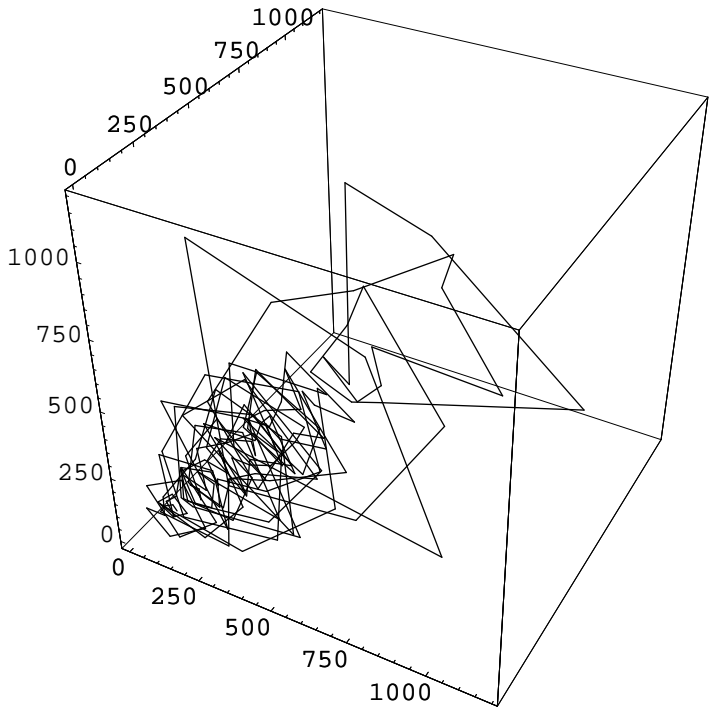}
\caption{The projection of the trajectory of the information flow
3-dimensional phase space.} \label{fig:atr}
\end{figure}

\begin{figure}
\includegraphics{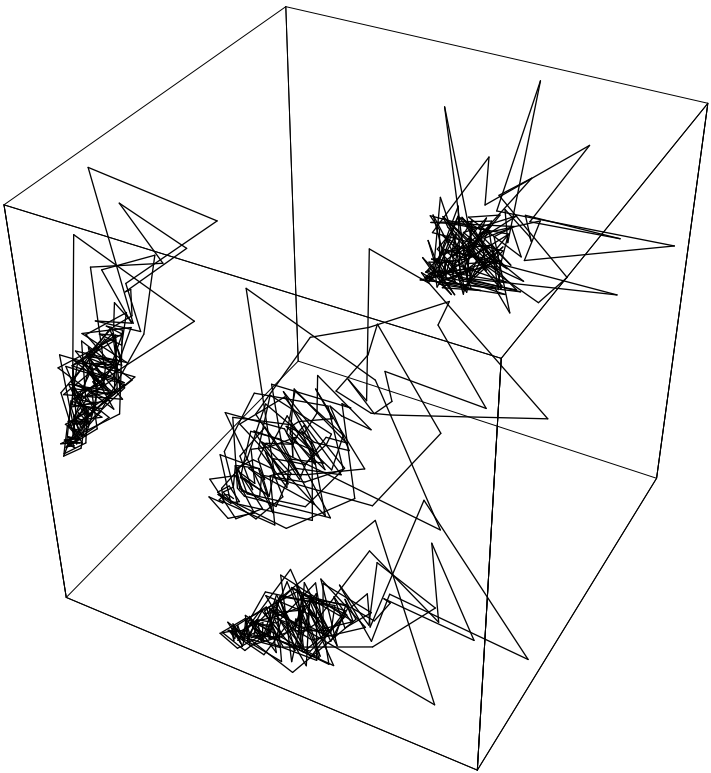}
\caption{The 2-dimensional projections of the trajectory
represented on Fig.(\ref{fig:atr}).} \label{fig:atr-pr}
\end{figure}

However, we do not know these independent parameters. Moreover,
due to the complexity of the system it is natural that these
unknown parameters, which are real dynamical degrees of freedom,
have very complicated relations with the parameters contained in
the network protocols.  Fortunately, these technique provide very
powerful methods to extract general information about behavior of
dynamical complex systems. For example, the obtained time
dependence of only one parameter, the protocol ID, shown on
Fig.(\ref{fig:ip}) is enough to reconstruct the trajectory of the
information flow in its phase space. The projection of the
trajectory on 3-dimensional space is shown on Fig.
(\ref{fig:atr}), and further three projections of the trajectory
on the 2-dimensional spaces are shown on Fig.(\ref{fig:atr-pr}).
One can see that the complete description of the network
information traffic in terms of small number of parameters is
possible and the the trajectory of information flow is rather well
localized and will be a subject for further analysis and pattern
recognition techniques.

Now we can obtain results for developing an approach to monitor
the network in real time and to prevent possible intrusions.

\section{Real Time Network Monitoring and Detection of Known Intrusions}

The direct consequences of the proposed approach for the network
traffic description provides the possibility of a real time
network monitoring and detection of all known network attacks.
This is because the complete information about network traffic at
any given point of the network is collected from tcpdump binary
output. All header parameters are converted into time dependant
numerical functions. Therefore, each packet for host-to-host
exchange corresponds to a point in the multidimensional parametric
phase space. The set of these points (the trajectory) completely
describes information transfer on the network. It is clear that
this representation provides not only the total description of the
network traffic at the given point but rather a powerful tool for
analysis in real time. Let us consider some possible scenarios for
the analysis.

Suppose we are looking for known network intrusions. The signature
of the intrusion is a special set of relations between the header
parameters. For example\cite{indet}, the signature for the attempt
to identify live hosts by those responding to the ACK scan
includes a source address, an ACK and SYN flags from TCP protocol,
a target address of internal network, sequences number, and source
and destination port numbers. The lone ACK flag set with identical
source and destination ports is the signature for the ACK scan,
since the lone ACK flag set should be found only as the final
transmission of the three-way handshake, an acknowledgement of
receiving data, or data that is transmitted where the entire
sending buffer has not been emptied. From this example one can see
that the intrusion signature could be easily formulated in terms
of logic rules and corresponding equations. Then, collecting the
header parameters (this is the initial phase of network
monitoring) and testing sets of them against the signatures
(functions in terms of the subset of the parameters) one can
filter out all known intrusions. Since we can collect any set of
the parameters and easily add any signature function, it provides
the way for a continuous upgrading of the intrusion detection
system (IDS) built on these principles.  In other words, such ID
system is universal and can be used to detect all possible network
intrusions by adding new filter functions or macros in the
existing testing program. It is very flexible and easily
upgradable.  The flexibility is important, however, and in
principle could be achieved even in existing ``traditional''
IDS's. What is out of scope of traditional approaches is the
minimization of possible false alarms and maximal optimized and
controlled sensitivity to intrusion signals. These features are an
intrinsic feature of our approach.

The important feature of the approach is the presentation of the
parameters in terms of time dependant functions. This gives the
opportunity to decrease the false alarm probability of the IDS as
best possible for the particular network. This can be done using
 sophisticated methods already developed for noise reduction in time series.
Moreover, representation of the protocol parameters as numerical
functions provides the opportunity for detailed mathematical
analysis and, as a consequence, for the optimization of the
signal-to-noise ratio using not only time series techniques but
also numerical methods for analysis of multi-dimensional
functions. The combination of these methods provides the best
possible way, in terms of accuracy of the algorithms and
reliability of the obtained information, for detection of known
intrusions in real time.

Also, the description of the information flow in terms of
numerical functions gives the opportunity to monitor the network
traffic for different time windows without missing information and
without overflowing storage facilities. One can suggest some
different ways to do it. One example is the use of a parallel
computer environment (such as low cost powerful Linux clusters)
for the simultaneous analysis of the decoded binary tcpdump
output. In this case the numerical functions of the header
parameters, being sent to different nodes of the cluster, will be
analyzed by each node using similar algorithms but different
scales for time averaging of signals (or functions). Thus each
node has separate time window and, therefore, is sensitive to the
network behavior in the particular range of time. Choosing time
averaging scales for the nodes from microseconds to weeks, for
example, one can trace and analyze the network traffic
independently and simultaneously in all these time windows. It is
worthwhile to remember that the optimal signal-to-noise ratio is
achieved for each time window independently providing the best
possible level of information traffic analysis for the whole
network. There are at least three obvious advantages for this
approach. The first one is the possibility to detect intrusions
developed on different time scales simultaneously and in real
time. The second one is the automatic decreasing of the noise from
a short time fluctuations for long time windows due to time
averaging. This provides detailed information analysis in each
time window without lost information and, at the same time,
discards all noise related information, drastically reducing the
amount of information at the storage facilities. The third
advantage is the possibility to use (in real time) output from
short time scale analyzed data as an additional information for
the long time scale analysis.

\begin{figure}
\includegraphics{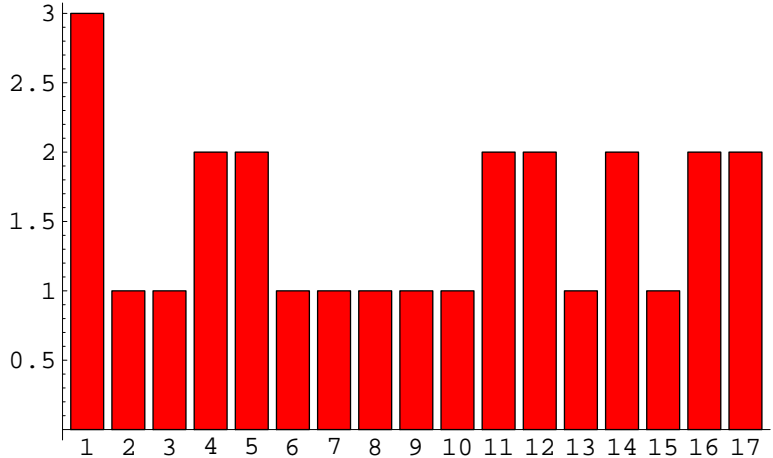}
\caption{Frequencies of the parameters used in signatures of
intrusions. For numbering rules see Table 1.} \label{fig:fr}
\end{figure}

Let us estimate how many parameters are used to describe
signatures of currently known intrusions. One analyzes the
comprehensive (but probably not a complete) list of known attacks:
smurf \cite{atk1}, fraggle \cite{atk2}, pingpong \cite{atk3}, ping
of death \cite{atk4}, IP Fragment overlap \cite{atk5}, BrKill
\cite{atk6}, land attack \cite{atk7}, SYN flood attack
\cite{atk8}, TCP session hijacking \cite{atk9}, out of band bug
\cite{atk10}, IP unaligned timestamp \cite{atk11}, bonk
\cite{atk12}, OOB data barf \cite{atk13} and vulnerability scans
(FIN and SYN \& FIN scanning) \cite{atk14}. The frequencies of the
parameters involved in signatures for these intrusions are shown
on Fig.(\ref{fig:fr}) and the numeration of the parameters is
explained in Table 1. One can see that the small number of
parameters that are used for signatures of intrusions. This fact
 simplifies the procedure of the analysis. \\

\begin{table}
\caption{The parameters involved in intrusion signatures as shown
on Fig.(\ref{fig:fr}).}
 \begin{tabular}{|c|c|c|c|c}
 \hline
 Number & Protocol & Parameter & Frequency \\
 \hline

1 &   IP & Destination IP Address  & 3 \\
2 &  IP & Source IP Address  &  1 \\
3 &  IP & Length  & 1 \\
4 & IP & More Fragment Flag &  2 \\
5 & IP & Don't Fragment Flag & 2\\
6 &  IP & Options & 1 \\
7 & TCP & Source Port & 1 \\
8 & TCP & Destination Port &  1 \\
9 & TCP & Urgent Flag & 1 \\
10 & TCP & RST Flag   & 1 \\
11 & TCP & ACK Flag   & 2 \\
12 & TCP & SYN Flag   & 2 \\
13 & TCP & FIN Flag   & 1 \\
14 & UDP & Destination Port  &  2 \\
15 & UDP & Source Port & 1 \\
16 & ICMP & Type & 2 \\
17 & ICMP & Code &  2 \\
\hline
 \end{tabular}
 \end{table}

\section{Detection of Unknown Intrusions}

The suggested approach could be considered as a powerful and
promising method for network monitoring and detection of known
network intrusions.  However, there is an even more important
feature: the ability to detect previously unknown attacks on a
network in a wide range of time scales. This ability is based on
the method of description of information exchange on networks in
terms of the exchange of packets provided by the header parameters
as described in terms of numerical functions of time (or a
trajectory in multi-dimensional phase space) as well as on
existing methods in theoretical physics for the analysis of
dynamics of complex systems. These methods lead to a very useful
 result for the dimensionality of the
information flow space. Since the number of parameters used in
packet header is high (of the order of hundreds), the practical
search for unknown (even abnormal) signal would be a difficult
problem if possible at all. Therefore, the reported result
\cite{gj1} on the small dimension of the parametric space of the
information flow is a crucial point for the practical approach for
the detection of unknown intrusions.

To build a real time intrusion detection system, which is capable
of detecting unknown attacks, we exploit the fact that one needs
to analyze only a small number of parameters. Furthermore, as it
is known from complex systems theory, the choice of the parameters
is not important unless they are sensitive to the system behavior.
The last statement needs to be explained in more detail.
Generally, hundreds different parameters could be encapsulated in
the packets headers. The question is which parameters we need to
choose for the right description of the information flow. One can
think, following the discussion in the previous section, that we
need to make our choice from the known quoted 17 parameters. It
may be a good guess. However, the number 17 is  bigger than the
dimension of the phase space which we are taking in mind, and it
could be a suspicion that hackers will invent new attacks with new
signature parameters which are not included in the set presented
in the previous section. The right answer on these remarks follows
from complex systems theory: for complete system description one
needs only the number of parameters equal to the phase space
dimension (more precisely, the smallest integer number which is
larger than fractal dimension of the phase space). It could be set
of any parameters which are sensitive to the system dynamics (and
the 17 discussed parameters could be good candidates). We do not
know, and do not suppose to know, the real set of parameters until
the theory of network information flow is developed or a reliable
model for information flow description is suggested. Nevertheless,
a method developed to study non-linear complex systems provides
tools to extract the essential information about the system from
the analysis of even a small partial set of the ``sensitive''
parameters. As an example, one can refer to the
Fig.(\ref{fig:atr}) which show the 3-dimensional projection of the
reconstructed trajectory from the time dependent behavior of only
one parameter (the protocol ID) shown on Fig.(\ref{fig:ip}). It
means that the complete description of the network information
flow could be obtained even from one ``sensitive'' parameter.

It should be noted that it is certainly not enough to analyze one
parameter for intrusion detection in realistic situation. In the
worst case, some analysis can be done using one parameter, but the
reliability, signal-to-noise ratio and time efficiency of the
analysis will be rather poor for it to be used as a search for an
unknown attack. However, the above extreme example demonstrates
the power of the method. The proper approach is to use as many
parameters as possible (up to the dimension of the phase space).
In that case one can reach the optimal sensitivity to an intrusion
signal and speed up the analysis procedure.

One of the ways to the realize  this approach is to use the
multi-window method discussed in the previous section with the
proper data analysis for each time scale. The method of analysis
is out of the scope of the current paper and will be reported
elsewhere. We will review only the general idea and the problems
related to this analysis.  To detect unknown attacks (unusual
network behavior) we use a deviation of signals from the normal
regular network behavior.  For these purposes one can use a
pattern recognition technique: as well to establish a normal
behavior patterns so as to measure a possible deviation from the
normal pattern. However, the pattern recognition problem is quite
difficult for this analysis. According to our knowledge, it is
technically impossible to achieve reliable efficiency in pattern
recognition for rather large, such as a 10 dimensional space. On
other hand, for our purposes, we cannot extract reliable
information from the analysis of one parameter. Therefore, one can
choose the optimal (compromised) solution to use pattern
recognition technique in the informational flow subspaces with low
dimensions. One can choose these subspaces as cross sections of
the total phase space defined by applying appropriate constraints
on some header parameters. In this case, we will have reasonable
ratio of signal-to-noise and will simplify and also improve
reliability for the pattern recognition technique. For pattern
recognition we suggest useing a 2-3 dimensional wavelet analysis
chosen on basis of detailed study of the information traffic on
the set of networks. The wavelet approach is promising because it
simultaneously drastically reduces the computational time and
memory requirements (which is very important for multidimensional
analysis) and because it can be used for additional effective
noise reduction technique.

\section{Conclusions}

We suggest a new approach for real time network monitoring which
is based on the application of complex systems theory for
information flow analysis on networks. The synthesis of the
network traffic description, in terms of numerical time dependant
functions, and methods of theoretical physics, for the study of
complex systems, provides not only a robust method for network
monitoring with detection of known intrusions, but it looks very
promising for developing real systems for detection of unknown
intrusions.

To apply innovative technology approaches that are measurably
effective
  against practical attacks it is necessary to detect and identify
  the attack on reconnaissance stage.
As an element for the future technology we presented the study of
the possibility of building an automatic intrusion detector
system, based on new methods of data analysis and pattern
recognition. The system will be able to help maintain a high level
of confidence in the protection of the networks.

We thank staff of Advanced Solutions Group for technical support.
 This work was supported by the DARPA Information Assurance and
 Survivability Program and is administered by the USAF Air Force
 Research Laboratory via grant F30602-99-2-0513, as modified.


\begin{thebibliography}{00}




\bibitem {cnet1}
A. R\'{e}ka, J. Hawoong and B. Albert-L\'{a}szl\'{o}, Nature, {\bf
406}, 378 (2000).

\bibitem {cnet2}
S. H. Strogatz, Nature, {\bf 410}, 268 (2000).

\bibitem {sig1}
L. Deri and S. Suin, Computer Networks {\bf 34}, 873 (2000).

\bibitem {st1}
P. A.  Porras and  A. Valdes, ``Live Traffic Analysis of TCP/IP
Gateways'', Internet Society Symposium on Network and Distributed
System Security, San Diego, California March 11-13, 1998

\bibitem {st2}
J. B. D. Cabrera, B. Ravichandram and R. K. Mehra,`` Statistical
Traffic Modeling for Network Intrusion Detection'', in:
Proceedings of the International Simposium on Modeling, Ananlysis
and Simulation of Computer and Telecommunication Systems, IEEE
(2000).

\bibitem {st3}
T. Huisinga {\it et al.}, arXiv:cond-mat/0102516 (2000).

\bibitem{spflow}
N. G. Duffield and M. Grossglauser, IEEE/ACM Transactions on
Networking, v. 9 no. 3 (June 2001) pp. 280-292.

\bibitem{timetr}
M. Ayedemir {\it et al.}, Computer Networks {\bf 36}, 169 (2001).

\bibitem {indet} S. Northcutt, J. Novak and D. McLachlan,
``Network Intrusion Detection, An Analyst's Handbook''  New Riders
Publishing, Indiapolis, IN, 2001.

\bibitem {tcpd} LBNL's Network Research Group , http://ee.lbl.gov/.

\bibitem {gj1} V. Gudkov and J. E. Johnson, submitted to Phys. Rev. Lett.

\bibitem {abar1} H. D. I. Abarbanel, R. Brown, J. J. Sidorowich and L. Sh. Tsimring,
Rev. Mod. Phys.  {\bf 65}, 1331 (1993).

\bibitem {chaos} J.-P. Ecmann and D. Ruelle,
Rev. Mod. Phys.  {\bf 57}, 617 (1985).

\bibitem {atk1}
http://www.insecure.org/sploits/routed.broadcast.ping.DOS.html;
http://advice.networkice.com/Advice/Exploits/IP/smurf/default.htm.

\bibitem {atk2}
http://advice.networkice.com/Advice/Intrusions/2000205/default.htm.

\bibitem {atk3}
http://advice.networkice.com/Advice/Exploits/IP/pingpong/default.htm.

\bibitem {atk4}
http://www.insecure.org/sploits/ping-o-death.html;
http://advice.networkice.com/Advice/Intrusions/2000012/default.htm.

\bibitem {atk5}
http://advice.networkice.com/Advice/Exploits/IP/fragments/default.htm;
http://www.insecure.org/sploits/linux.fragmentation.teardrop.html.

\bibitem {atk6}
http://www.rootshell.com/archive-j457nxiqi3gq59dv/199810/brkill.c;
http://deep.ee.siue.edu/br/brkill/brkill.html.

\bibitem {atk7}
http://www.insecure.org/sploits/land.ip.DOS.html;
http://advice.networkice.com/advice/exploits/tcp/land/default.htm.

\bibitem {atk8}
http://advice.networkice.com/advice/exploits/tcp/syn\%5Fflood/default.htm.

\bibitem {atk9}
http://www.microsoft.com/technet/network/intern.asp.

\bibitem {atk10}
http://ideval.ll.mit.edu/Links/attackDB.html.

\bibitem {atk11}
http://www.ee.siue.edu/\~rwalden/networking/ipopt.html;
http://www.securityfocus.com/bid/1173.html.

\bibitem {atk12}
http://www.insecure.org/sploits/95.NT.fragmentation.bonk.html.

\bibitem {atk13}
http://www.insecure.org/sploits/windows.OOB.DOS.html.

\bibitem {atk14}
http://www.whitehats.com/IDS/27;
http://www.networkice.com/advice/Underground/Hacking/Methods/
Technical/Port\_Scan/default.htm;
http://www.geocities.com/the\_darkdream/port\_scanning\_unscanned.htm.



\end{thebibliography}
\end{document}